\documentclass[pre,showpacs,preprint]{revtex4}

\usepackage{amsmath}

\usepackage{graphicx}
\usepackage{dcolumn}
\usepackage{bm}

\begin{document}

\title{Rheological effects in the linear response and spontaneous fluctuations  of a sheared granular gas }
\author{J. Javier Brey, P. Maynar, and M.I. Garc\'{\i}a de Soria}
\affiliation{F\'{\i}sica Te\'{o}rica, Universidad de Sevilla,
Apartado de Correos 1065, E-41080, Sevilla, Spain}
\date{\today }

\begin{abstract}
The decay of a small homogeneous perturbation of the temperature of a dilute granular gas in the steady uniform shear flow state is investigated. Using kinetic theory based on the inelastic Boltzmann equation, a closed equation for the decay of the perturbation is derived. The equation involves the generalized shear viscosity of the gas in the time-dependent shear flow state, and therefore it predicts relevant rheological effects beyond the quasi-elastic limit. A good agreement is found when comparing the theory with molecular dynamics simulation results. Moreover, the Onsager postulate on the regression of fluctuations  is fulfilled.

\end{abstract}

\pacs{45.70.-n,05.20.Dd, 05.60.-k,51.10.+y}

\maketitle

\section{Introduction}
\label{s1}
The study of the linear response of an equilibrium system to a small perturbation either of the hydrodynamic fields or of an external field, is a formally exact method of non-equilibrium statistical mechanics, that has led to a significant advance in the theory of simple atomic liquids \cite{Fo75}.
Nevertheless, little is known  with the same degree of generality and exactness about the linear response of fluids that are in non-equilibrium steady states. Granular gases provide an appropriate context in which to address this as well as many other general questions about irreversibility \cite{Du00,Go03}. This is because granular gases are intrinsic non-equilibrium systems, but also because they exhibit steady states that are macroscopically simple, as compared with those of molecular fluids.

The objective of this work is to provide a case in which the theory can be developed in a controlled way and to compare its predictions with  molecular dynamics simulations. The linear response of a dilute granular gas in a reference homogeneous steady state with constant shear rate is studied, for arbitrary values of the latter.  Although this state can not be implemented in real experiments, it has the advantage of being easily obtained by means of particle simulation methods. The considered perturbation is homogenous and affects the internal energy, or granular temperature, of the system but not the macroscopic velocity field.  Rheological effects  are known to be relevant in the reference state when the shear rate, measured in units of the inverse of the average time between collisions, is finite \cite{JyR88,BRyM97,MGSyB99}.  These effects manifest themselves in the strong dependence of the components of the pressure tensor on the shear rate, beyond the proportionality relation assumed by the Navier-Stokes approximation. A question addressed here is how this rheology translates into the linear response of the system to the perturbation.

 A key ingredient in the linear response theory for molecular fluids is the Onsager postulate \cite{On31}, asserting that the average regression of thermal fluctuations in equilibrium follows the macroscopic linear law of relaxation. This property is now quite well understood \cite{Fo75}, but little is known about its validity in far from equilibrium states.  This issue will be investigated here for a granular gas under steady shear flow.

The remaining of this paper is organized as follows. In Sec. \ref{s2}, the equations characterizing a dilute granular gas under uniform shear flow are presented, and the existence of a steady solution is indicated.  A small homogeneous perturbation of the temperature of the steady state is considered, and the linear equation describing its time evolution is derived.  The scaling verified by the distribution function of the system is indicated and its consequences on several macroscopic properties of the relaxing system are discussed. This leads to a formal equation for the decay of the temperature. This equation is transformed into a closed hydrodynamic form in Sec. \ref{s3}, by employing known properties of the reference state. Then, the theoretical prediction is compared with molecular dynamic results, and a good agreement is obtained, although some systematic discrepancy shows up for very strong inelasticity. Moreover, it is shown that the linear perturbation decays with the same rate as the temperature time-correlation function in the reference state, i.e. the Onsager postulate is verified. Finally, Sec. \ref{s4} is used to present a short summary of the results obtained in the paper.

\section{Decay of a temperature perturbation}
\label{s2}
The focus here is on the decay of a small homogeneous perturbation of the temperature in a granular gas in the steady uniform shear flow (USF) state. Although explicit calculations will be restricted to the low density limit, their extension to higher densities may be possible. Consider first a granular gas in the time-dependent USF state. At the macroscopic level, this state is characterized by a uniform number density $n$, a time-dependent uniform granular temperature $T(t)$, and a constant velocity field ${\bm u}$ with linear profile \cite{Haff90,BRyM97,JyR88,SGyN96},
\begin{equation}
\label{2.1}
u_{i}({\bm r})= \sum_{j} a_{ij} r_{j}, \quad a_{ij} \equiv a \delta_{ix} \delta_{jy},
\end{equation}
where $\delta_{ij}$ is the Kroneker delta function and $a$ the constant shear rate. For this state, the mass conservation is identically verified, the conservation of momentum requires that the $xy$ component of the pressure tensor $P_{xy}$ is uniform, and the energy balance equation reads \cite{BRyM97}
\begin{equation}
\label{2.2}
\frac{\partial T(t)}{dt}= - \frac{2a}{nd}\, P_{xy}(t)- \zeta(t) T(t).
\end{equation}
 Here, $d$ is the dimension, $2$ or $3$, of the system, and $\zeta(t)$ is the cooling rate due to the energy dissipation in collisions. The above relation admits a steady form,
\begin{equation}
\label{2.3}
\frac{2a}{nd}\, P_{xy,s}=-\zeta_{s} T_{s}.
\end{equation}
Equations (\ref{2.2}) and (\ref{2.3}) are exact consequences of the balance equations, but they become meaningful only after some constitutive relations for $P_{xy}$ and $\zeta(t)$ are given. These relations   express the cooling rate and the pressure tensor as functionals of  the hydrodynamic fields, $n$, ${\bm u}$, and $T$. The particular case of a low density granular gas composed of smooth inelastic hard spheres ($d=3$) or disks ($d=2$) of mass $m$ and diameter $\sigma$ will be considered.  For states that are homogeneous in the Lagragian frame of reference for the velocity field given in Eq.\ (\ref{2.1}), the one particle distribution of the gas, $f({\bm V},t)$, where ${\bm V}({\bm r}) ={\bm v}-{\bm u}({\bm r})$ is the peculiar vecity of the particle, obeys the inelastic Boltzmann equation
\begin{equation}
\label{2.4}
\left(\frac{\partial}{\partial t} -a V_{y} \frac{\partial}{\partial V_{x}} \right) f({\bm V},t)= J[{\bm V}|f(t)],
\end{equation}
where $J[{\bm V}|f(t)]$ is the Boltzmann collison operator for smooth inelastic hard spheres or disks \cite{GyS95}. The shear  stress $P_{xy}$ and the cooling rate $\zeta(t)$ in Eq.\ (\ref{2.2}) are functionals of the one-particle distribution function,
\begin{equation}
\label{2.5}
P_{xy}(t)= m \int d{\bm V}\, V_{x}V_{y} f({\bm V},t),
\end{equation}
\begin{equation}
\label{2.6}
\zeta(t)=\frac{(1-\alpha^{2}) \pi^{(d-1)/2} m \sigma^{d-1}}{4 \Gamma \left( \frac{d+3}{2} \right) n T(t)d}\, \int d{\bm V}_{1} \int d{\bm V}_{2}\,  V_{12}^{3} f({\bm V}_{1},t) f({\bm V}_{2},t),
\end{equation}
with ${\bm V}_{12} \equiv {\bm V}_{1} -{\bm V}_{2}$. A sufficient condition for the existence of constitutive relations and hence of an hydrodynamic description, is that the distribution function of the system be of the kind named normal \cite{RydL77,DyB06}, and meaning that all the time dependence in it occurs through the temperature, i.e., it has the form
\begin{equation}
\label{2.7}
f({\bm V},t)= n \left[ v_{0}(t) \right]^{-d} \chi ({\bm c}, a^{*}),
\end{equation}
where
\begin{equation}
\label{2.8}
v_{0}(t) \equiv \left[ \frac{2T(t)}{m} \right]^{1/2}
\end{equation}
is a thermal velocity,
\begin{equation}
\label{2.9}
{\bm c} \equiv \frac{\bm V}{v_{0}(t)}\, ,
\end{equation}
and
\begin{equation}
\label{2.10}
a^{*} \equiv \frac{a \lambda}{v_{0}(t)}\, ,
\end{equation}
with $\lambda \equiv (n \sigma^{d-1})^{-1}$ being a quantity proportional to the mean free path of the gas. The distribution function of the steady USF state, is obtained by particularizing the above expression for the steady temperature, namely
\begin{equation}
\label{2.11}
\chi ({\bm c},a^{*}) \rightarrow \chi ({\bm c}, a_{s}^{*}),
\end{equation}
\begin{equation}
a^{*}_{s}\equiv \frac{a \lambda}{v_{0,s}}, \quad v_{0,s} \equiv \left( \frac{2T_{s}}{m} \right)^{1/2}.
\end{equation}

 The idea that for a wide class of initial conditions the steady USF state is reached, after a short initial transient period,  through a hydrodynamic evolution of the system has been already used. In particular, the way in which the normal solution is approached has been investigated \cite{AyS07,AyS12}.
Suppose that the system is in the steady USF state and then it is submitted at $t=0$ to an instantaneous and homogeneous small perturbation of the temperature (or internal energy), keeping constant the value of the shear rate $a$. It is assumed that the subsequent time evolution of the temperature is described by hydrodynamics, so that it will obey the linearization of Eq.\ (\ref{2.2}) around the steady USF state. Such a linearization reads
\begin{equation}
\label{2.12}
\frac{\partial \delta T(t)}{\partial t}= -\frac{2a}{nd}\, \delta P_{xy}(t)-T_{s} \delta \zeta(t) - \zeta_{s} \delta T(t),
\end{equation}
where $\delta T(t) \equiv T(t) - T_{s}$, and $\delta P_{xy}$ and $\delta \zeta$  are the linear deviations of the shear stress and the cooling rate, respectively, from their steady values. In the hydrodynamic regime, Eq. (\ref{2.5}) can be rewritten as
\begin{equation}
\label{2.13}
P_{xy}(t)= \frac{1}{2}\, n m v_{0}^{2}(t) P^{*}_{xy} (a^{*}),
\end{equation}
with the reduced shear stress defined as
\begin{equation}
\label{2.14}
P^{*}_{xy}(a^{*}) \equiv 2 \int d{\bm c}\, c_{x}c_{y} \chi({\bm c},a^{*}).
\end{equation}
Then, a simple calculation gives
\begin{equation}
\label{2.15}
\delta P_{xy}(t)=2 P_{xy,s} \frac{\delta v_{0}(t)}{v_{0,s}}- \frac{1}{2}\, n m v_{0,s}^{2} a_{s}^{*} \left( \frac{ \partial P^{*}_{xy}(a^{*})}{\partial a^{*}} \right)_{a^{*}=a^{*}_{s}} \frac{\delta v_{0}(t)}{v_{0,s}}\, .
\end{equation}
It is convenient to introduce a generalized shear viscosity $\eta^{*}(a^{*})$ through
\begin{equation}
\label{2.16}
P_{xy}^{*}(a^{*})=-\eta^{*}(a^{*}) a^{*}.
\end{equation}
In the Navier-Stokes (NS) approximation, $\eta^{*}$ does not depend on $a^{*}$ by definition.  Its dependence outside that limit is referred to as the rheological effects on the viscosity of the system. Use of Eq.\ (\ref{2.16}) into Eq. (\ref{2.15}) yields
\begin{equation}
\label{2.17}
\delta P_{xy}(t)= \frac{1}{2} \left[P_{xy,s}+\frac{1}{2}\, n m v_{0,s}^{2} a_{s}^{*2} \left( \frac{\partial \eta^{*}(a^{*})}{\partial a^{*}} \right)_{a^{*}=a^{*}_{s}}\right] \frac{\delta T(t)}{T_{s}}\, ,
\end{equation}
where it has been used that
\begin{equation}
\label{2.18}
\frac{\delta v_{0}(t)}{v_{0,s}} = \frac{1}{2}  \frac{\delta T(t)}{T_{s}}\, .
\end{equation}
Next, the second term on the right hand side of Eq.\ (\ref{2.12}) will be considered. Substitution of the scaled form of the normal solution, Eq.\ (\ref{2.7}) into Eq.\, (\ref{2.6}) yields
\begin{equation}
\label{2.19}
\zeta (t)= \frac{v_{0}(t)}{\lambda}\, \zeta^{*}(a^{*}),
\end{equation}
with the dimensionless cooling rate $\zeta^{*}$ given by
\begin{equation}
\label{2.20}
\zeta^{*}(a^{*}) = \frac{(1-\alpha^{2}) \pi^{\frac{d-1}{2}}}{2 \Gamma \left( \frac{d+3}{2} \right)d}\, \int  d{\bm c}_{1} \int d{\bm c}_{2}\, c_{12}^{3} \chi ({\bm c}_{1},a^{*}) \chi ({\bm c}_{2},a^{*})\, .
\end{equation}
Therefore,
\begin{equation}
\label{2.21}
\delta \zeta(t)= \frac{v_{0,s}}{2 \lambda} \left[ \zeta^{*}(a^{*}_{s})-a^{*}_{s} \left( \frac{\partial \zeta^{*}(a^{*})}{\partial a^{*}} \right)_{a^{*}=a^{*}_{s}} \right] \frac{\delta T(t)}{T_{s}}.
\end{equation}
Now Eqs. (\ref{2.17}) and (\ref{2.21}) will be substitute into Eq.\ (\ref{2.12}). A dimensionless representation will be employed. The relative deviation of the temperature
\begin{equation}
\label{2.22}
\theta (t) \equiv \frac{\delta T (t)}{T_{s}},
\end{equation}
is introduced, as well as a dimensionless time scale $s$ defined by
\begin{equation}
\label{2.23}
s \equiv \frac{v_{0,s}}{\lambda}\, t.
\end{equation}
The time $s$ is proportional to the average number of collisions per particle in the time interval between $0$ and $t$. In this way, Eq. (\ref{2.12}) takes the form
\begin{equation}
\label{2.24}
\frac{\partial \theta(s)}{\partial s} = - \left[ \zeta^{*}(a^{*}_{s})+ \frac{a_{s}^{*3}}{d} \left( \frac{\partial \eta^{*}(a^{*})}{\partial a^{*}} \right)_{a^{*}=a^{*}_{s}}- \frac{a^{*}_{s}}{2} \left( \frac{\partial \zeta^{*}(a^{*})}{\partial a^{*}} \right)_{a^{*}=a^{*}_{s}} \right] \theta (s).
\end{equation}
Upon deriving this equation, the steady relationship given in Eq.\ (\ref{2.3}) has been taken into account.
It follows that a small homogeneous perturbation of the temperature of a granular gas under steady uniform shear flow decays exponentially in time,
\begin{equation}
\label{2.25}
\theta(s) = e^{-\gamma s} \theta (0),
\end{equation}
with the relaxation rate
\begin{equation}
\label{2.26}
\gamma =  \zeta^{*}(a^{*}_{s})+ \frac{a_{s}^{*3}}{d} \left( \frac{\partial \eta^{*}(a^{*})}{\partial a^{*}} \right)_{a^{*}=a^{*}_{s}}- \frac{a^{*}_{s}}{2} \left( \frac{\partial \zeta^{*}(a^{*})}{\partial a^{*}} \right)_{a^{*}=a^{*}_{s}}\,  .
\end{equation}
Note that, because of the constant linear relation between the dimensionless time scale $s$ and the original one $t$, the decay of the perturbation in the latter scale is also exponential.

\section{Molecular Dynamics simulations}
\label{s3}

Equation (\ref{2.25}) follows directly, without any additional approximation, from the hypothesis that the linear relaxation of the system after the perturbation is described by a normal distribution having the scaling form (\ref{2.7}). But, to actually compute the decay rate $\gamma$, the dependence of both the cooling rate and the shear viscosity on the shear rate are needed. This would require to know the explicit form of the scaled distribution $\chi ({\bm c}, a^{*})$.  In the NS approximation, the shear viscosity is independent of the shear rate as already mentioned and, for a dilute granular gas \cite{BDKyS98,ByC01}, the dependence of the cooling rate on the shear rate has been shown to be negligible in the linear approximation. Therefore, to NS order the rate $\gamma$ can be accurately approximated by the reduced cooling rate $\zeta^{*}_{0}$ computed to zeroth order in the gradients, i.e. the one of the homogeneous cooling state particularized for the density and temperature of the steady USF state. An accurate expression for this property is \cite{GyS95,vNyE98},
\begin{equation}
\label{3.1}
\zeta^{*}_{0} \approx \frac{\sqrt{2} \pi^{(d-1)/2}(1-\alpha^{2})}{\Gamma \left( d/2 \right) d}\, \left( 1+\frac{3c_{2}}{16} \right),
\end{equation}
\begin{equation}
\label{3.1a}
c_{2}(\alpha) = \frac{16 (1-\alpha)(1-2 \alpha^{2})}{9+24d +(8d-41)\alpha + 30 \alpha^{2} -30 \alpha^{3}}\, .
\end{equation}
In the steady USF state, the inelasticity, measured through the coefficient of normal restitution $\alpha$, and the shear rate $a^{*}$ are not independent, but they are related by Eq.\ (\ref{2.3}). As a consequence, the restriction to small gradients required by the NS approximation, also implies limitation to small inelasticity, or to the quasielastic limit $\alpha \rightarrow 1$.  The analysis was consistently extended to the next order in $a$, Burnett order, in ref. \cite{SGyN96}. The general case of arbitrary shear rate (and inelasticity), has been studied by considering a  simple kinetic model in which the inelastic Boltzmann collision operator is replaced by a single relaxation towards local equilibrium term, plus another term describing the energy dissipation due to the inelasticity of collisions \cite{BRyM97,SyA05}. Also, the Grad approximation to the Boltzmann equation has been used \cite{Ga02}. Quite interestingly, all the above studies lead to expressions for the generalized shear viscosity that are hardly distinguishable over the complete interval $0 \leq \alpha < 1$ and that are in very good agreement \cite{BRyM97,AyS05} with simulation results obtained by means of the direct Monte Carlo simulation method \cite{Bi94}. Moreover, another general feature of all the mentioned previous studies is that they do not lead to any shear dependence of the cooling rate of the steady USF state. Although we are not aware of any direct test of this theoretical prediction by means of numerical simulations, the agreement observed for the steady temperature provides a strong support for it, since the expression of $T_{s}$  involves both the shear viscosity and the cooling rate (see Eq. (\ref{2.3})).

An accurate expression for the generalized shear viscosity of the time-dependent uniform shear flow state has been obtained in refs.  \cite{AyS07,AyS12}. In our notation, it reads,
\begin{equation}
\label{3.1b}
\eta^{*}(a^{*}) = \frac{(d+2) \Gamma \left( d/2 \right)}{2 \sqrt{2} \pi^{\frac{d-1}{2}} \beta \left[ 1+2G(a^{*}) \right]^{2} \left[2+ H(a^{*})\right]}\, ,
\end{equation}
where
\begin{equation}
\label{3.1c}
G(a^{*})= \frac{2}{3}\,  \sinh^{2} \left\{ \frac{1}{6} \cosh^{-1} \left[ 1+ \frac{27}{32 d \pi^{d-1}} \left( \frac{(d+2) \Gamma \left( d/2 \right) a^{*}}{\beta} \right)^{2} \right] \right\},
\end{equation}
\begin{equation}
\label{3.1d}
H(a^{*}) = \frac{\left[ \frac{(d+2)(1- \alpha^{2})}{4 \beta d} - 2 G(a^{*}) \right]\left[ 1- 6 G(a^{*}) \right]}{\left[ 1+ 6 G(a^{*}) \right]^{2}},
\end{equation}
\begin{equation}
\label{3.1e}
\beta= \frac{1+\alpha}{2}\, .
\end{equation}
For the sake of completeness, also the expression of the steady reducedd shear  rate is written down,
\begin{equation}
\label{3.1f}
a^{*}_{s}= \frac{4 \sqrt{2} \pi^{\frac{d-1}{2}}}{(d+2) \Gamma \left( d/2 \right) } \left[ \frac{d(d+2)(1-\alpha^{2})}{8 \beta d} \right]^{1/2} \left[ \beta + \frac{d+2}{4d} (1- \alpha^{2} )  \right] .
\end{equation}

Finally, $\partial \zeta^{*}(a^{*})/ \partial a^{*}$ will be also neglected in Eq.\
(\ref{2.26}). Again, this is consistent with the previous model studies, but no numerical test of it has been carried out up to now, to the best of our knowledge. In summary, we approximate the expression for the rate of the decay of the temperature perturbation in Eq.\ (\ref{2.26}) by
\begin{equation}
\label{3.2}
\gamma \approx \zeta^{*}_{0}+  \left( \frac{a^{*3}}{d} \frac{\partial \eta^{*}(a^{*})}{\partial a^{*}} \right)_{a^{*}=a^{*}_{s}},
\end{equation}
with $\zeta^{*}_{0}$ given by Eq.\ (\ref{3.1}) and the shear viscosity $\eta^{*}(a^{*})$ by Eq.\ (\ref{3.1b}). For $d=3$,  Eq.\ (\ref{3.2}) is the same as obtained by particularizing the expressions derived in ref. \cite{Ga06} by means of an extension of the Chapman-Enskog expansion to inhomogeneous steady states \cite{Lu06},  applied to the simple relaxation kinetic model equation mentioned above.

In order to test the accuracy  of Eqs. (\ref{2.25}) and (\ref{3.2}), we have carried out event driven molecular dynamics (MD) simulations \cite{AyT87} of a system of $N=2000$ hard disks using Lees-Edwards periodic boundary conditions \cite{LyE72}. A square box was used and the density of the system was $n=0.02 \sigma^{-2}$, which is low enough as to expect the low density limit analysis to be accurate. The value of the shear rate was in all cases $a= 6.32 \times 10^{-3} \left( T(0)/m \right)^{1/2} \sigma^{-1}$, where $T(0)$ is the initial temperature. Finally, the value of the coefficient of normal restitution was varied in the simulations.

\begin{figure}
\includegraphics[scale=0.7,angle=0]{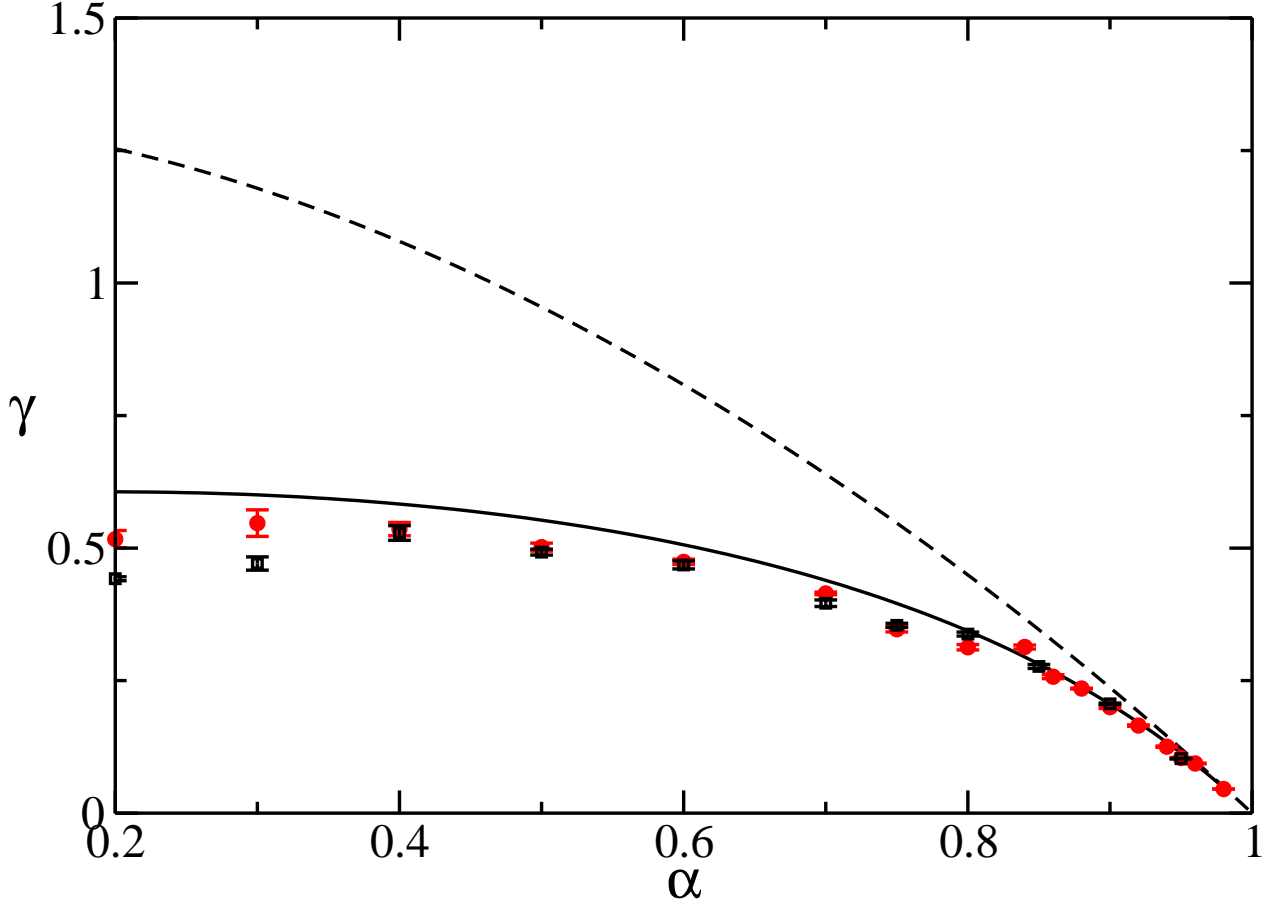}
\caption{(Color online) Dimensionless decay rate $\gamma$ of an small homogeneous perturbation of the temperature of a dilute granular in the steady uniform shear flow state as a function of the coefficient of normal restitution $\alpha$. The solid line is the theoretical prediction given by Eq.\ (\protect{\ref{3.2}}), the dashed line is the dimensionless reduced shear rate $\zeta^{*}_{0}$, the (black) squares are the MD simulation results from the decay of an initial perturbation, and the (red) circles have been obtained from the decay of the energy time-correlation function in the steady state. } \label{f1}
\end{figure}

Once the steady USF state was reached in the simulations, the small temperature perturbation was implemented as follows. The $y$-component of the velocity of all the particles was multiplied by the constant factor $1.2$, i.e. the velocities were modified as  $v_{y} \rightarrow 1.2 v_{y}$. Note that this scaling of the velocity does not modify the macroscopic velocity field, keeping in particular a vanishing average value of its $y$-component. Therefore, this perturbation belongs to the class of perturbations considered in Sec.\ \ref{s2}. In the simulations, an exponential decay of the temperature after the perturbation was observed, accordingly with the theoretical result in Eq.\ (\ref{2.25}).  The single  exponential shape of the decay guarantees that underlaying equation for the relaxation is linear and hence the validity of the linear approximation. In addition, it has been checked that equivalent results are obtained for smaller values of the perturbation of the velocity. Then, by means of a numerical fitting, the decay rate $\gamma$ was measured. The comparison of the simulation results and the
theoretical prediction given in Eq.\ (\ref{3.2}) is shown in Fig. \ref{f1}. A fairly good agreement is observed except for very strong inelasticity ($\alpha$ very small). The discrepancy is probably due to the system approaching the instability point of the steady state \cite{WyB86}. To stress the relevance of the rheological effects on the rate, the Navier-Stokes rate, given by the dimensionless cooling rate $\zeta^{*}_{0}$, has also been plotted.

An alternative way of measuring the decay rate $\gamma$ follows if the Onsager hypothesis on the regression of the equilibrium fluctuations \cite{On31} is extended and assumed to apply also to the internal energy of the steady USF state of a granular gas.  Then, the decay of the internal energy time-correlation function in the steady state is also $\gamma$ \cite{BGyM12}. The values obtained in this way have also been included in Fig.\ \ref{f1}. They agree with the results got from the decay of the initial perturbation, within the statistical uncertainties, with some discrepancy in the very inelastic limit. Again, this can be due to the proximity of the instability of the steady state, which can affect in a different way the time fluctuations and the linear response.

\section{Final comments}
\label{s4}
In this paper, the decay of an infinitesimal homogeneous perturbation of the temperature of a dilute granular gas in the steady USF state has been investigated, by using scaling properties of the normal solution of the Boltzmann equation. It has been shown that relevant rheological effects show up beyond the quasi-elastic limit. They tend to slow down the increase of the relaxation rate as the restitution coefficient decreases. Moreover, the rheology in the energy decay is associated to the shear rate dependence of the shear viscosity, while the dependence of the cooling rate, if it exits, it plays a negligible role. This confirms the results derived previously on the basis of model kinetic equations.

Also, it has been verified that the energy time-correlation function in the
steady USF state decays with the same rate as the energy perturbation, then obeying a hydrodynamic description. This generalizes to a far from trivial situation the Osager hypothesis on the regression of equilibrium fluctuations.

The natural question to be addressed in the near future is whether the conclusions reached here also apply to global properties of other systems in non-equilibrium steady states, both in granular gases  and molecular fluids. If this were the case, it would indicate the existence of some generalization to non-equilibrium of the Onsager principle. This would be consistent with the structure of the fluctuating Navier-Stokes equations recently derived for dilute granular gases \cite{BMyG11}. Also, a few related studies concerning the validity of  fluctuation-dissipation relations in driven dilute granular gases have already been carried out
\cite{PByL02,PByV07,MGyT09}.

\section{Acknowledgements}

This research was supported by the Ministerio de Educaci\'{o}n y Ciencia (Spain)
through Grant No. FIS2011-24460 (partially financed by FEDER funds).

\end{document}